\documentclass[twocolumn,showpacs,preprintnumbers,amsmath,amssymb]{revtex4}

\usepackage{graphicx}
\usepackage{dcolumn}
\usepackage{bm}

\newcommand{\eqb}{\begin{equation}}
\newcommand{\eqe}{\end{equation}}
\newcommand{\pd}{\partial}

\newcommand{\eab}{\begin{eqnarray}}
\newcommand{\eae}{\end{eqnarray}}

\newcommand{\e}{\mbox{e}}

\newcommand{\La}{\Lambda}

\begin{document}
\title{Analytical approach to nonperturbative Yang-Mills thermodynamics}
\author{Ralf Hofmann}\vspace{0.3cm}
\affiliation{Institut f\"ur Theoretische Physik,
Universit\"at Heidelberg,
Philosophenweg 16,
69120~Heidelberg,
Germany}

\begin{abstract}
An analytical and inductive approach to hot SU(N) Yang-Mills dynamics is developed. 
For N=2,3 pressure and energy density are pointwise compared with lattice data. 


\end{abstract} 

\maketitle

\indent {\sl Introduction.} An analytical grasp of the thermodynamics of 
SU(3) gauge theory is needed to interprete 
ultrarelativistic heavy ion collisions. 
Thermal perturbation theory (TPT) by itself 
is unable to provide a framework for this. 
Nonperturbative lattice input is needed for a prediction of the thermal pressure 
\cite{Schroeder2001-2003}. Lattice experiments are 
a nonperturbative, first-principle approach. Although numerical in nature, 
they have boosted our 
knowledge about relevant degrees 
of freedom and their condensation. It is by now an established result that condensed 
center vortices are the relevant degrees of freedom for confinement 
at temperature $T$ below $T_c$ \cite{CenterVortex}. 
The purpose of this Letter is to develop a 
nonperturbative approach to thermal SU($N$) thermodynamics 
based on thermal quasiparticle excitations (QPEs) and the condensation of 
the solitonic objects caloron, magnetic monopole and center vortex (pointlike in 
4D, 3D, and 2D \cite{calexp}). The latter processes 
successively reduces the asymptotic gauge symmetry SU(N) 
according to the pattern SU(N)$\to$U(1)$^{\tiny\mbox{N-1}}$$\to$Z$_{\tiny\mbox{N}}\to$\,nothing. 
The first two symmetry reductions are accompanied by a 2$^{\mbox{\tiny{nd}}}$ (possibly occuring close to the Planck mass) 
and a very mild 1$^{\mbox{\tiny{nd}}}$ order thermal phase transition, respectively. 
Order parameters for these transitions are the masses of 
gauge bosons associated with coset generators and the Cartan subalgebra, respectively.  
The final transition towards spontaneously broken, local Z$_{\tiny\mbox{N}}$ symmetry 
is 1$^{\mbox{\tiny{st}}}$ order. The order parameter is the energy density $\rho$. 
A single mass scale $\La_E$ or $\La_M$ or $\La_C$, determined by a boundary condition, introduces a gauge 
invariant scale separation in all phases.\\  
\indent {\sl Analysis.} 
At temperatures 
very much higher than the dynamic 
scale $\La_{\tiny\mbox{YM}}(\mbox{N})$ TPT, when applied to the calculation of 
thermodynamic potentials, shows convergence 
in its perturbative orders.\\ 
{\sl {\bf Electric phase.}} At a critical temperature 
$T^{P}_c$, much larger 
than $\La_{\tiny\mbox{YM}}(\mbox{N})$, a thermal 2$^{\mbox{\tiny{nd}}}$ 
order phase transition occurs at which the system condenses 
a composite, adjoint Higgs field $\phi$ reducing the gauge symmetry 
maximally, SU(N)$\to$ U(1)$^{\tiny\mbox{N-1}}$ (lowering $T$ across the transition, 
the order parameter gauge boson mass continuously increases starting 
from a very small screening mass ). 
Relevant degrees of freedom driving this condensation are 
calorons. At high temperature 
an isolated caloron is a pointlike object with a nontrivial holonomy and topological 
charge one in 4D Euclidean spacetime. A caloron possesses N 
Bogomoln'yi--Prasad--Sommerfield (BPS) magnetic monopole constituents \cite{Nahm1984,KraanVanBaal1998} related by center
transformations. Its action is $S_{\mbox{\tiny cal}}=8\pi^2/e^2$ 
\cite{KraanVanBaal1998}, where $e$ denotes the 
gauge coupling. The caloron is a BPS saturated 
solution to the SU(N) Yang-Mills equation of motion , 
its 3D energy density vanishes.  It is a coherent thermal 
state in 4D.  The picture that caloron 
condensation causes the growth of gauge boson masses 
with decreasing $T$ is self-consistent if 
$S_{\mbox{\tiny cal}}$ rapidly decreases 
below $T^P_c$ by a large rise of the 
gauge coupling $e$. This indeed happens, 
see Fig.\,1a. An indirect argument in favor of 
caloron {\sl condensation} relies on the 
lattice result of a vanishing 4D topological 
charge density $\chi$ for $T>T_c$, implying that 
calorons do not populate the ground state as 
small-size objects. The caloron radius and $e$ 
are strictly monotonously related. We conclude that above 
$T_c$ $e$ is large and the caloron action small. 
After caloron condensation has taken place, the proposed effective action is\\ 
\vspace{-0.4cm} 
\eab
\label{actE}
\hspace{-0.5cm}S_E\hspace{-0.2cm}&=&\hspace{-0.2cm}\int_0^{1/T}\hspace{-0.5cm}
d\tau\hspace{-0.1cm}\int \hspace{-0.1cm}d^3x\,\left(\frac{1}{2}\,\mbox{tr}
\,G_{\mu\nu}G_{\mu\nu}+\mbox{tr}\,{\cal D}_\mu\phi{\cal D}_\mu\phi+
V_E(\phi)\right)
\eae
where $V_E$ denotes the potential for the 
field $\phi$. The covariant derivative is 
defined as ${\cal D}_\mu\phi=\pd_\mu+ie[\phi,A_\mu]$ and the field 
strength as $G_{\mu\nu}=G^a_{\mu\nu}t^a$, where 
$G^a_{\mu\nu}=\pd_\mu A^a_\nu-\pd_\nu A^a_\mu-ef^{abc}A^b_\mu A^c_\nu$ 
and $\mbox{tr}_{\tiny\mbox{N}}\,t^a t^b=1/2\delta^{ab}$. In a suitable gauge the field $\phi$ describes the coherent 
ground state of the thermal system. 
As a consequence, its gauge invariant modulus $|\phi|$ 
is independent of spatial coordinates. A dependence of $|\phi|$ 
on $\tau$ may only enter through an 
adiabatically slow $\tau$ dependence of the 
temperature $T$. In the absence of gauge fields in 
(\ref{actE}) the 3D energy density of $\phi$ is exactly zero since 
it is composed of BPS saturated calorons. This condition is 
satisfied for the configuration $\phi$ 
if and only if it is BPS saturated itself. In other words: we search a 
potential $V_E$ such that (i) a BPS saturated and spatially homogeneous 
solution to the classical equation of motion, derived from (\ref{actE}) 
in the absence of gauge fields, exist, (ii) the modulus $|\phi|$ 
of this configuration is independent of $\tau$, and 
(iii) quantum and thermal fluctuations of $\phi$ are negligible. 
We proceed by discussing the case of even N. The other case 
is more complicated, an analysis is given in \cite{PRL3}. Here results 
for N=3 are quoted in square brackets if appropriate. 
We work in a gauge where $\phi\equiv\mbox{diag}(\tilde{\phi}_1,
\tilde{\phi}_2,\cdots,\tilde{\phi}_{{\tiny\mbox{N}}/2})$ 
is SU(2) block diagonal \cite{oddN}\\ 
\vspace{-0.4cm} 
\eab
\label{Vn1/2}
\hspace{-0.5cm}v_E&\equiv&i\Lambda_E^3\,\mbox{diag}
(\lambda_1\tilde{\phi}_1/|\tilde{\phi}_1|^2,\cdots,
\lambda_1\tilde{\phi}_{{\tiny\mbox{N}}/2}/|\tilde{\phi}_{{\tiny\mbox{N}}/2}|^2)\,.
\eae
The SU(2) modulus is defined as $|\tilde{\phi}_l|^2\equiv 1/2\,\mbox{tr}_2\,\tilde{\phi_l}^2$ 
$(l=1,\cdots,\mbox{N}/2)$. The gauge invariant potential $V_E$ is given as 
$V_E(\phi)=\Lambda_E^6\,\mbox{tr}_{\tiny\mbox{N}}\,(\phi^2)^{-1}$. 
The periodic solution to the BPS equation 
$\pd_{\tau}\phi=v_E$ ($V_E\equiv \mbox{tr}_{\tiny\mbox{N}} v_E^2$) with 
minimal potential and maximal SU($N$) symmetry breaking 
is in SU(2) decomposition given as \\ 
\vspace{-0.4cm}
\eab
\label{persoln}
\hspace{-0.5cm}\tilde{\phi}_l(\tau)&=&\sqrt{\Lambda_E^3/(2\pi T l)}\,\lambda_3
\exp(-2\pi i T l\lambda_1\tau)\,. 
\eae
Obviously, $\phi$ is a traceless, 
hermitian matrix. In unitary gauge, 
$\phi=\mbox{diag}(\phi_1,\phi_2,\cdots,\phi_N)$, 
there is no $\tau$ dependence 
(no $\tau$ dependence occurs in $\mbox{tr}_N\,\phi^2$). 
From (\ref{Vn1/2}) and (\ref{persoln}) we obtain the following 
ratios: $\pd^2_{|\tilde{\phi}_l|}V_E/T^2=12\pi^2\,l^2$ and 
$\pd^2_{|\tilde{\phi}_l|}V_E/|\tilde{\phi}_l|^2=3l^3\lambda_E^3$ where 
the dimensionless temperature is defined as $\lambda_E\equiv 2\pi T/\La_E\gg 1$. 
As a consequence, thermal fluctuations of $\phi$ are negligibly small and 
quantum fluctuations do not exist (the mass of $\tilde{\phi}_l$ excitations is larger 
than the compositeness scale $|\tilde{\phi}_l|$). A pure-gauge 
solution to the equation of 
motion ${\cal D}_\mu G_{\mu\nu}=2ie[\phi,{\cal D}_\nu\phi]$ 
in the background of (\ref{persoln}) is 
$A^{bg}_\mu=(\pi/e)\,T \delta_{\mu 0}\,\mbox{diag}
(\lambda_1, 2\,\lambda_1,\cdots, \mbox{N}/2\,\lambda_1)$. 
We have ${\cal D}_\mu\phi=0$ for $A^{bg}_\mu$ entering in ${\cal D}_\mu$. 
As a consequence, the $\phi$ kinetic term in the action 
(\ref{actE}) vanishes on $\phi$ and $A^{bg}_\mu$, 
the ground-state energy density is lifted 
to $V_E(\phi)=\pi/2\,\Lambda_E^3 T \mbox{N(N+2)}$ 
[$V_E(\phi)=4\pi\,\Lambda_E^3 T$] by caloron interaction. 
The field $\phi$ back-reacts by emitting and 
absorbing the thermal quasiparticle excitations (QPEs) described by fluctuations $a_\mu$. 
This makes sense since $\phi$ is composed of coherent 
thermal states - calorons. Substituting $A_\mu=A^{bg}_\mu+a_\mu$ 
into the action (\ref{actE}), its form is retained as an action for 
the fluctuations $a_\mu$ (the $\phi$ kinetic term induces mass 
terms for $a_\mu$) only in unitary 
gauge, where $A^{bg}_\mu=0$. QPEs are related to 
fluctuations $a_\mu$ only in this gauge. Unitary gauge is reached by a 
gauge transformation $A_\mu\to \Omega^\dagger A_\mu\Omega+i/e(\pd_\mu\Omega^\dagger)\Omega$ 
($\Omega\equiv e^{i\theta}$) involving a {\sl nonperiodic} $\theta$ 
with SU(2) decomposition $\theta_l=-\pi \lambda_1 Tl\tau$. Because the matter 
field $\phi$ does not fluctuate 
the physics is inert under such a gauge transformation: Hosotani's 
mechanism \cite{Hosotani1983} does not take place. Out of the N$^2$-1 
independent modes N-1 modes reside in the Cartan subalgebra 
and remain massless on tree level (TLM modes). 
In unitary gauge the N(N-1) independent off-diagonal modes,  
defined w.r.t. the SU(N) generators 
$t^{IJ}_{rs}=1/2\,(\delta_r^I\delta_s^J+\delta_s^I\delta_r^J)$ and 
$\bar{t}^{IJ}_{rs}=-i/2\,(\delta_r^I\delta_s^J-\delta_s^I\delta_r^J)$ 
($I=1,\cdots,$N) ($J>I$), are heavy on tree level 
(TLH modes). Their mass spectrum is obtained from the 
$\phi$ kinetic term in (\ref{actE}) as $m_a^2=-2e^2\,\mbox{tr}\,[\phi,t^{a}][\phi,t^{a}]$ 
($t^{a}=\{t^{IJ},\bar{t}^{IJ}\}$). We have $m_{IJ}^2=e^2(\phi_I-\phi_J)^2$ 
for both $t^{IJ}$ and $\bar{t}^{IJ}$. 
Since TLH and TLM modes are QPEs radiative corrections are {\sl local} in unitary gauge, unitarity is trivially satisfied. 
A radiative correction $\Delta m_{IJ}^2$ to $m_{IJ}^2$ and a radiatively induced mass-squared for a TLM mode 
$m^2_{TLM}$ are determined 
by a {\sl single} 4-gauge boson tadpole with a TLH or a TLM mode and a 
TLH mode running in the loop, respectively. Since a loop integral is cut off at the compositeness 
scale $|\phi|$, the expansion of a radiative correction is obtained as 
$\Delta m_{IJ}^2=|\phi|^2/(8\pi^2)(c_{-1}^{\tiny\mbox{TLH}}e^2+c_0^{TLH}e^0+c_1^{\tiny\mbox{TLH}}e^{-2}+\cdots)$ and 
$m^2_{\tiny\mbox{TLM}}=|\phi|^2/(8\pi^2)(c_{0}^{\tiny\mbox{TLM}}e^0+c_1^{\tiny\mbox{TLM}}e^{-2}+\cdots)$. 
For N=2 we have $c_0^{\tiny\mbox{TLM}}=1/8$. Radiative corrections $\Delta V_E$ 
to the potential $V_E$ are induced by one-loop and two-loop bubbles. In the latter case two one-loop bubbles are 
connected by a 4-gauge boson vertex. $\Delta V_E$ is tiny. 
To lowest oder in $1/e^2$ we have $\Delta V_E/V_E<(\mbox{N}^2-1)/(16\pi^2)(|\phi|/\La_E)^6\ll 1$; 
see below. We proceed by considering tree-level masses only. 
A useful quantity is $a_k=c_k a$ (mass/$T$ of a TLH mode) 
where $a\equiv e\sqrt{\Lambda_E^3/(2\pi T^3)}$ 
(or $a\equiv g\sqrt{\Lambda_M^3/(2\pi T^3)}$, see below). The value 
of the dimensionless constant $c_k$ ($k=1,\cdots,\mbox{N(N-1)}$) 
derives from the TLH mass spectrum. TLH modes are 
thermal QPEs - they acquire mass and {\sl structure} by interacting with 
the ground state of the system. 
Thus the thermodynamic relation 
$\rho_E=T\frac{dP_E}{dT}-P_E$ between the total energy density 
$\rho_E(T)$ and pressure $P_E(T)$ does not 
follow from the partition function associated with 
the action (\ref{actE}) for the fluctuations $a_\mu$. 
This problem is resolved by imposing the condition 
of minimal thermodynamic self-consistency $\pd_a P=0$ \cite{Gorenstein1995}. 
An evolution equation follows as\\ 
\vspace{-0.4cm}
\eab
\label{eeq}
\pd_a \lambda_E=-\frac{24\,\lambda_E^4\,a}{(2\pi)^6 
\mbox{N(N+2)}}\sum_{k=1}^{\tiny\mbox{N(N-1)}}c_k^2 D(a_k)\,,
\eae
where $D(a)\equiv \int_0^{\infty} dx\,
x^2/(\sqrt{x^2+a^2}(\exp(\sqrt{x^2+a^2})-1))$. 
The right-hand side of (\ref{eeq}) is negative definite: the 
function $\lambda_E(a)$ can be inverted to $a(\lambda_E)$. 
As a consequence, the gauge coupling is obtained as 
$e(\lambda_E)=a(\lambda_E)\lambda_E^{3/2}/(2\pi)$. 
To compute $e(\lambda_E)$ numerically, subject to 
the initial condition $a(\lambda_E^P=1000)=0$, we have devised a 
Mathematica program. The result is shown in Fig.\,1a.
\begin{figure}
\begin{center}
\leavevmode
\leavevmode
\vspace{3.5cm}
\includegraphics{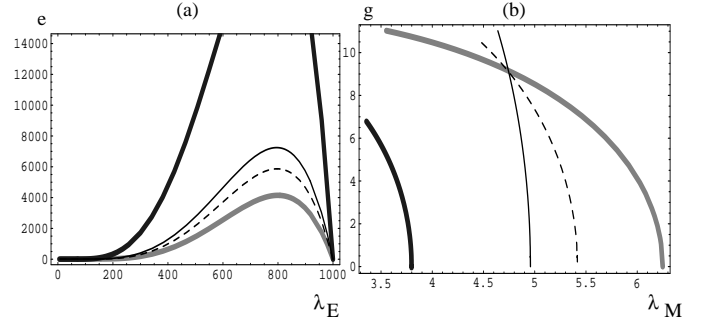}
\end{center}
\caption{The evolution of the gauge 
coupling in the electric (a) and magnetic phase (b) for 
N=2 (thick grey lines), N=3 (thick black lines), N=4 (dashed lines), and N=10 
(thin solid lines). The initial conditions in (b) are obtained by 
matching the pressure at the phase boundary. 
In both cases the gauge couplings 
diverge logarithmically, $e,g\sim -\log(\lambda_{E,M}-\lambda^c_{E,M})$, at 
$\lambda^c_E=\{9.92,4.79,8.6,7.87\}$ and 
$\lambda^c_M=\{3.55,3.36,4.48,4.64\}$, respectively. This 
is not captured by Mathematica numerics.}      
\end{figure}
We have checked numerically that the value of 
$\lambda^c_E$ does not depend on the initial 
condition $a(\lambda_E^P)=0$ within a wide 
range of $\lambda_E^P$ values. The plateau 
values of $e$ are $e\sim\{17.15,8,14,12\}$ for N=2,3,4,10, respectively. We conclude that the 
caloron action $S_{\mbox{\tiny cal}}=8\pi^2/e^2$ is very small throughout 
the entire electric phase. The assumed condensation 
of calorons, driving the condensation of $\phi$, 
is self-consistent by a large 
rise of $e$ across the phase boundary. There is a regime in $\lambda_E$ 
where $e$ is large but much smaller than its value close to the critical values 
$\lambda_E^c$ and $\lambda_E^P$. Calorons are of finite size but still large enough not to generate a 
topological susceptibility on state-of-the-art lattices. Constituent 
BPS magnetic monopoles become visible 
\cite{KraanVanBaal1998}. Calorons start to 
scatter elastically (SU(2) calorons of the same embedding in SU(N) go through each other), scatter 
inelastically (calorons of different embedding approach), 
and annihilate (caloron meets its anticaloron). No monopoles, single monopoles, and 
monopole-antimonopole pairs arise, respectively. Monopoles are 
slowed down by surrounding calorons. A pair separates into 
stable particles connected by a magnetic flux line, a long-lived dipole forms. 
If a single monopole is produced then it is unstable (magnetic charge is conserved!) 
unless it connects with its antimonopole, produced in 
an independent collision. There are N species of monopoles equal to the maximal number 
of monopole constituents a free-of-magnetic-charge SU(N) caloron can have. 
How do we understand this in the effective electric theory? 
In the gauge, where $\phi$ is SU(2) block diagonal, the temporal winding 
of the $l^{\tiny\mbox{th}}$ block is complemented by spatial winding at isolated points in 3D space. 
By a large, $\tau$ 
dependent gauge transformation the monopole's asymptotic SU(2) 
Higgs field is rotated to spatial constancy and into the direction given 
by the temporal winding of the unperturbed block $\tilde{\phi}_l$. Its 
Dirac string rotates in Euclidean time. Since this monopole is stable, 
a correlated antimonopole must exist. We arrive at a dipole rotating 
about its center of mass at an 
angular frequency $2\pi l T$. This rotation is an 
artifact of our gauge choice. Rotating the dipole to unitary gauge by the 
above gauge function $\theta_l$, we arrive 
at a (quasi)static dipole. Recall that the fluctuations $a_\mu$ are 
thermal QPEs only in unitary gauge. There are isolated 
coincidence points (CPs) in time where the lower right (upper left) corner $\varphi_l$ 
($\varphi_{l+1}$) of the $l^{\tiny\mbox{th}}$ ($(l+1)^{\tiny\mbox{th}}$) SU(2) block (now ($l=1,\cdots,\mbox{N}/2-1$))
together with the number zero of its 
right-hand (left-hand) neighbour are proportional to the generator $\lambda_3$. 
Coincidence also happens between the first and last diagonal entry in $\phi$. 
At a CP, spatial winding may take place at isolated points in 3D space. 
Monopoles of this type are unstable particles. Coincidence also happens between nonadjacent 
SU(2) blocks, see Fig.\,2. The associated 
monopoles are, however, not independent. We obtain N/2+N/2=N monopole species - 
in accord with the above counting [The field $\phi$ winds with winding number one 
in each of the independent SU(2) subalgebras for half the time. The matching 
happens at the CPs $\tau_{CP}=0,1/(2T)$ where $\phi(\tau_{CP}\pm 0)$ 
add up to an element of the third, dependent SU(2) algebra. At these CPs we may have 
an unstable monopole.]. Monopoles have resonant excitations and are coherent thermal states 
in a thermal environment \cite{ForgasVolkov2003}.\\   
{\sl {\bf Magnetic phase.}} 
\begin{figure}
\begin{center}
\leavevmode
\leavevmode
\vspace{3cm}
\includegraphics{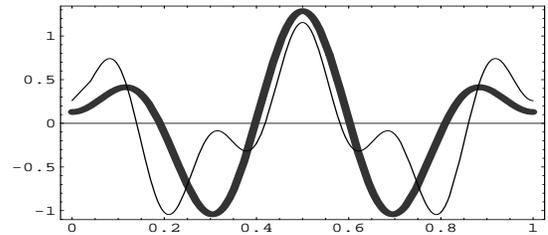}
\end{center}
\caption{Sum of $\varphi_2$ and $\varphi_3$ (bold curve) 
and sum of $\varphi_2$ and $\varphi_5$ as a function of $\tau$. 
A zero in each curve is associated with 
an unstable monopole or its unstable antimonopole.}      
\end{figure}
The mass $M_l$ of a stable (unstable) BPS monopole is given 
as $M_l\sim 8(1/\sqrt{2})\pi |\tilde{\phi_l}|/e$. It vanishes at $\lambda^c_E$. 
As a consequence, $M_l$ and the 
monopole action $M_l/T$ vanish. TLH modes decouple at $\lambda^c_E$. Magnetic monopoles condense in a 
2$^{\mbox{\tiny{nd}}}$ order phase transition (radiative corrections in the electric 
phase lift this to a very weak 
1$^{\mbox{\tiny{st}}}$ order transition). The order parameter is the TLM mass. 
It is continuous across the transition. U$^{\tiny\mbox{N-1}}$ is spontaneously broken by the 
condensates of stable monopoles 
described by the complex fields $\phi_k$ ($k=1,\cdots,\mbox{N}/2\,[2]$). These fields are 
winding in time with integer winding number. The center symmetry 
Z$_N$ is unbroken. It is represented by a local permutation symmetry 
$\phi_k\to\phi_{k+j}$ $(j(\vec{x})=1,\cdots,\mbox{N}-1$; for $k>\mbox{N}/2$ it maps zero onto zero) 
under which 
the ground state is inert. The potential for the complex field 
$\phi_k$ is defined as $V_M^{(k)}\equiv \bar{v}_M^{(k)}v_M^{(k)}$, where 
$v_M^{(k)}=i\La_M^3/\phi_k$. The ground state of the system 
possesses an energy density $1/2\,V_M=(\pi/8)\,T\Lambda_M^3 \mbox{N(N+2)}$ [$1/2\,V_M=\pi\,T\Lambda_M^3$] 
(canonical definition of $\phi$-kinetic
term). It is constructed in close 
analogy to the electric phase. Again, it is assured that there are no thermal and quantum fluctuations of the 
fields $\phi_k$. In unitary gauge, $\phi_k=|\phi_k|$, we define QPEs as thermal fluctuations of the 
Abelian gauge bosons. These QPEs are all massive away from the phase boundary. Their mass spectrum is $m_k=g|\phi_k|$, 
where $g$ denotes the magnetic gauge coupling. At the electric--magnetic phase boundary, 
we have $g=2\pi/e=0$. There are no radiative corrections to the gauge boson masses in the magnetic 
phase. Minimal thermodynamic self-consistency 
implies an evolution equation for $g$: $\pd_a\lambda_M=-96\lambda_M^4 a
/((2\pi)^6 \mbox{N(N+2)})\sum_{k=1}^{\tiny{\mbox{N/2}}}c_k^2 D(a_k)$ 
[$\pd_a\lambda_M\hspace{-0.15cm}=\hspace{-0.15cm}-12\lambda_M^4 a D(a)
/(2\pi)^6$]. The evolution of 
$g$ with temperature is shown in Fig.\,1b. 
Monopole condensation is driven by a 
large rise in the gauge coupling $g$ across the phase boundary. 
The size of a BPS magnetic monopole is given 
by a constant $C^{-1}\propto g$ \cite{PrasadSommerfield1975}. According to Fig.\,1b the size of 
a magnetic monopole increases. Isolated dipoles and unstable 
single monopoles `feel' each other close to the boundary 
to the center phase, where $g$ diverges logarithmically. The former form stable strings 
of dipoles. The latter form stable
dipoles first and then stable strings of dipoles. Strings of dipoles are interpreted 
as fat center vortices \cite{CenterVortex}. Monopole constituents of a vortex are coherent 
thermal states and so is a vortex.\\ 
{\sl {\bf Center phase.}}
Although a center vortex and an 
Abrikosov--Nielsen--Oleson (ANO) vortex are not 
the same solitonic object, the Euclidean action and the energy $E_{\tiny\mbox{vor}}$ of the former can be 
estimated by the latter. For a typical ANO 
vortex (length $\sim$ width) they are given as 
$S_{\tiny\mbox{ANO}}\sim 1/g^{2}$ and $E_{\tiny\mbox{vor}}\propto 1/g^{2}$ \cite{NielsenOlesen1973}. 
Since $g$ diverges 
logarithmically at $\lambda^c_M$  center vortices condense and 
Abelian gauge bosons decouple. The equation of state (EOS) at $\lambda^c_M$ is $\rho=-P$. 
The condensed phase is described by complex and 3D local fields $\Phi_n,\Phi_{\bar n}$ $n=1,\cdots \mbox{N}$. 
These fields are, up to a dimensionful (and $T$ dependent) normalization, the condensates of 
spherical, spatial Wilson loops of diameter 
comparable to the core-sizes of the respective center vortices. 
The (local!) Z$_{\tiny\mbox{N}}$ symmetry acts as $\Phi_n\to\e^{2\pi i p/\tiny\mbox{N}}\Phi_{n+p},
\Phi_{\bar n}\to\e^{2\pi i p/\tiny\mbox{N}}\Phi_{n+p}$ 
($p=p(\vec{x})=1,\cdots,\mbox{N}$). Thus it is spontaneously broken.\hspace{-0.2cm}The thermodynamics of the 
fields $\Phi_n$ ($n=1,\cdots,\mbox{N}$) is determined by $Z_{\tiny\mbox{N}}$ symmetric 
potentials $V_C^{(n)}\hspace{-0.12cm}\equiv\hspace{-0.12cm}
\bar{v}_C^{(n)}v_C^{(n)}$ where 
$v_C^{(n)}=i(\La_C^3/\Phi_{n}-\Phi_{n}^{\tiny{\mbox{N-1}}}/\La_C^{\mbox{N-3}})$. The N 
degenerate minima of the 
potential $V_C^{(n)}$at $|\Phi_n|=\La_C$ are {\sl exactly at zero}. 
At $\lambda_M^c$ a jump of the order parameter $\rho$ to $\rho=0$ takes place. 
The latent heat $\epsilon_C$ is given as $\epsilon_C=(\mbox{N(N+2)}/16)\lambda_M^c \La_M^4$ 
[$\epsilon_C=(1/2)\lambda_M^c\La_M^4$]. 
The pressure $P$ in the center phase starts at the negative value $-2\epsilon_C$ and relaxes rapidly to zero. 
If a matching to the magnetic phase would take place in 
thermal equilibrium no QPEs would exist below $\lambda_M^c$ since the potential is well 
approximated by its pure pole term, and  
the ussual calculation for $\pd^2_{|\Phi_n|}V_C^{(n)}$ applies. 
This situation holds for $|\Phi_n|$ sufficiently smaller than $\La_C$ or for very large N. 
The solutions to the BPS equations would then be identical to the ones in the magnetic phase. 
Deviations from thermal equilibrium at the center transition manifest themselves 
in a $\tau$ dependence of $|\Phi_{n}|$, see \cite{Hofmann2000}. Since $Z_{\tiny\mbox{N}}$ is a local symmetry 
no domain walls are generated in the transition. Heavy `glueballs' are produced by a rapid re-heating, see \cite{PRL2} 
for the interpretation of these `hadrons'. 
 \\   
{\sl {\bf Matching the phases.}} $P$ and $T$ are continuous across 
a thermal phase boundary. In the electric and magnetic phase the ratio of gauge and $\phi$ kinetic terms 
defines the mass of the gauge-field spectrum at a given $T$. A re-definition of this ratio re-defines the scales 
$\La_E$ and $\La_M$, respectively. Using the canonical definition (also for the center phase) 
and disregarding a polarization mismatch for 
Abelian gauge bosons at the electric--magnetic phase boundary, this relates the 
scales $\Lambda_E,\Lambda_M,\Lambda_C$ as 
$\Lambda_E=(1/4)^{1/3}\Lambda_M=(1/4)^{1/3}\Lambda_C$ 
[$\Lambda_E=(1/2)^{1/3}\Lambda_M=(1/3)^{1/3}\Lambda_C$] on tree-level.\\  
{\sl {\bf Energy density $\rho$ and pressure $P$.}} 
In Fig.\,3 the $T$ dependence of $\rho$ and $P$ is shown. The results 
for other thermodynamic potentials and for the critical exponents of the 
perturbative-electric and electric--magnetic transition will 
be published elsewhere. 
The value of $\rho$ at $\lambda_M^c$ is a measure of the latent heat released in the 
1$^{\mbox{\tiny{st}}}$ order-like transition following the onset of 
center-vortex condensation. 
\\ 
{\sl Deviation from thermal 
equilibrium (DTE) at $\lambda^M_C$:} A measure for this deviation 
is the relative deviation of the full potential $V_C$ from 
its pure pole term evaluated on 
the lowest winding mode taken at $\lambda_M=\lambda^c_M$. For $N=2,3,4,10$ 
we respectively obtain: 70\%, 41\%, 23\%, and 1\%. 
The large DTE for N=2 may explain the apparent 2$^{\mbox{\tiny{nd}}}$ 
order deconfinement transition seen on
the lattice. It may be an 
artifact of imposing thermal equilibrium in the lattice simulation. 
For N=3 the DTE is still large. It is, however, conspicuous that a high-statistics determination of 
$P$, not relying on the method of 
$\beta$ integration, obtains a negative pressure 
slightly above $T_c$ \cite{Deng1988}. 
The $\beta$-integration method may be biased by a 
prescribed integration constant. For N=4 and in particular for 
N=10 the situation is much better for lattice simulations. It is extremly likely that 
future lattice simulations with N$>$3 will unambiguously identify negative pressure 
close to $\lambda^c_M$ (our confidence regarding the truth of this prediction 
derives from the results in \cite{PRL2}). For N=2,3,4,10 we predict {\sl negative} pressure of 
magnitude as given in Fig.\,3. \\  
{\sl {\bf Comparison with the lattice.}}
\begin{figure}
\begin{center}
\leavevmode
\leavevmode
\vspace{5.5cm}
\includegraphics{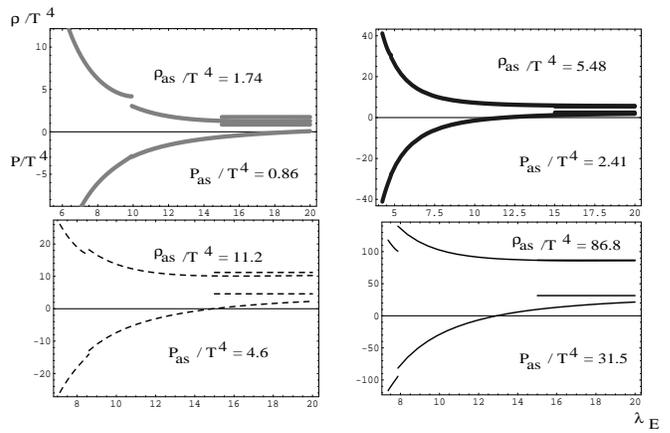}
\end{center}
\caption{The normalized energy density $\rho/T^4$ and the normalized pressure 
$P/T^4$ as functions of $\lambda_E$. The horizontal 
lines indicate the respective asymptotic value 
(taken at $\lambda_E=100$). The discontinuities are due to the jump 
in gauge-boson polarizations across the electric--magnetic transition, 
this effect disappears beyond tree-level (Abelian gauge bosons 
in the electric phase are then massive). Notice the brevity of the 
magnetic phase.}      
\end{figure}
For comparison with lattice data we take 
$T_c$ to correspond to $\lambda_M^c$.  
For $N=2$ we obtain: $(\rho_{\tiny\mbox{lat}}/\rho)(T=1.535;4.878;12.925\,T_c)=0.31;1.26;1.29$ 
\cite{EJKLLNS1987,EKSM1982,EKSM1981} ($18^3\times5;10^3\times3;10^3\times5$ lattices), 
$(P_{lat}/P)(T=4.878\,T_c)=1.26$ \cite{EKSM1982} 
($10^3\times3$ lattice). For $T=12.195\,T_c$ $P_{\tiny\mbox{lat}}$ is calculated using 
$\Delta_{\tiny\mbox{lat}}=\rho_{\tiny\mbox{lat}}-3P_{\tiny\mbox{lat}}$ and $\rho_{lat}$. For $N=3$ we obtain: 
$\rho_{lat}/\rho(T=5\,T_c)=0.88$, $P_{lat}/P(T=5\,T_c)=0.74$ 
\cite{BoydEngelsKarschLaermannLegelandLuetgemeierPeterson1996} ($32^3\times8$ lattice). Deviations may be explained 
by the tree-level treatment of gauge boson masses in the 
electric phase and a mismatch with the lattice scale \cite{IC}.\\  
{\sl Summary}. 
We have developed a nonperturbative, inductive approach to thermal SU(N) Yang-Mills theory. The numerical 
results for pressure and energy density are promising. A more precise computation will 
aim at a precise match of the lattice scale and the scale $\La_C$ including radiative corrections in the electric phase. 
This predicts the critical temperature $T^c_P$ in the UV regime (possibly close to the Planck mass). 
The detection of negative pressure at $T$ close to $T_c$ has far-reaching consequences for our 
understanding of accelerated cosmological expansion and the creation of a lepton asymmetry \cite{PRL2}. 
QCD thermodynamics will be described by an extension of the approach which includes dynamical, 
fundamental fermions, see \cite{PRL2} for an analogy. We believe that the mechanisms 
of gauge-symmetry breakdown presented here have a lot to 
say about electroweak symmetry breaking {\sl without} the conventional prediction of 
Higgs-particle excitations.     
\\ {\sl Acknowledgments:} It is a pleasure to thank 
Philippe de Forcrand, Daniel Litim, and Zurab Tavartkiladze 
for useful discussions. The author would like to thank 
Valentine Zakharov for many wonderful discussions focussing on the physics of 
magnetic monopoles in non-Abelian gauge theories. The hospitality and financial support extended 
to the author by CERN's theory division are thankfully acknowledged.

\bibliographystyle{prsty}

\begin{thebibliography}{10}
\bibitem{calexp}
Whenever we speak of the condensation of a 4D, 3D, or 
2D particle the condensation of the 
associated antiparticle is also understood
\bibitem{oddN} For odd N an SU(2) block on the diagonal is 
replaced by an SU(3) block, Pauli matrices 
$\lambda_i$ ($i=1,2,3$) are replaced as 
$\lambda_i\to\bar{\lambda}_i$ where $\bar{\lambda}_3$ 
either $\mbox{diag}(0,1,-1)$ or $\mbox{diag}(1,0,-1)$, 
for N$>$3 an average over these replacements must be performed; not 
performing this average makes no difference 
for the potential and the spectrum in the electric and magnetic phases 
but changes the physics in the center phase
\bibitem{IC} The scale $\La_C$ does not correspond to the absolute 
lattice scale $\La_L$, at large $T$ this mismatch becomes 
less important than at low $T$; to perform a precise 
numerical match is beyond the scope of this Letter and should be 
done using high-precision numerics
\end{thebibliography}

\end{document}